\documentclass[conference]{IEEEtran}
\usepackage{url}
\usepackage{ifpdf}
\usepackage{multirow}

\usepackage{color}

\usepackage{cite}
\ifCLASSINFOpdf
   \usepackage[pdftex]{graphicx}
\else
   \usepackage[dvips]{graphicx}
   \graphicspath{{../eps/}}
\fi
\usepackage[cmex10]{amsmath}
\usepackage{array}
\usepackage{mdwmath}
\usepackage{mdwtab}
\usepackage[font=footnotesize]{subfig}
\newcommand{\bps}{\tilde{b}_{ps}^{(k)}}
\newcommand{\aps}{\tilde{a}_{ps}^{(k)}}
\begin{document}
\title{Analytical Modeling of Saturation Throughput in Power Save Mode of an IEEE 802.11 Infrastructure WLAN}


\author{\IEEEauthorblockN{Pranav Agrawal\IEEEauthorrefmark{1},
Anurag Kumar\IEEEauthorrefmark{2},
Joy Kuri\IEEEauthorrefmark{1}, 
Manoj K. Panda\IEEEauthorrefmark{2}
\IEEEauthorblockA{\IEEEauthorrefmark{1}Centre for Electronics Design and Technology }
\IEEEauthorblockA{\IEEEauthorrefmark{2}Electrical Communication Engineering}
Indian Institute of Science, Bangalore, India - 560 012}}

% \author{}


\maketitle

\begin{abstract}
%\boldmath

We consider a single station (STA) in the Power Save Mode (PSM) of an IEEE 802.11  infrastructure WLAN. This STA is assumed to be carrying uplink and downlink traffic via the access point (AP). We assume that the transmission queues of the AP and the STA are saturated, i.e., the AP and the STA  always have at least one packet to send. For this scenario, it is observed that uplink and downlink throughputs achieved are different. The reason behind the difference is the long term attempt rates of the STA and the AP due to the PSM protocol. In this paper we first obtain the the long term attempt rates of the STA and the AP and using these, we obtain the saturation throughputs of the AP and the STA. We provide a validation of analytical results using the NS-2 simulator.
 \end{abstract}

% \fi
\IEEEpeerreviewmaketitle

\section{Introduction}

In the normal mode of operation, also called the Continuously Active Mode (CAM), an STA always keeps its radio on, so it can receive and transmit at any time. This mode of operation is energy inefficient since STAs draw current even when they are idling. To save power during the period when there is less or no network activity, WiFi cards are provided with controls through which they can be turned off. To leverage this facility, the IEEE 802.11 standard has a feature using which STAs can turn off their radio without losing packets. This mode of operation is called as the Power Save Mode (PSM). In this mode, an STA can be in any one of the two state, \emph{active state} and \emph{sleep state}. In the active state the radio of the card is turned on, so it can receive as well as transmit, where as in the sleep state, the radio of the card is turned off, hence in this state the STA can neither receive nor transmit. Generally PSM is efficient only when there is activity of very short duration during which the STA uploads and downloads bursts of packets. In this paper, we focus on this burst of activity, and assuming that during this activity both the STA and the AP are saturated. We observed that the throughput obtained by the AP is higher than that of the STA. This difference of the throughput is attributed to the different long term attempt rate of the STA and the AP, and attempt rate of the AP being higher.

\emph{Contribution:}
We consider a scenario in which a single STA is associated with the Access Point (AP). The STA is considered to be in PSM and carrying uplink and downlink traffic via the AP. We assume that the AP and the STA are saturated, i.e., immediately after the transmission of a packet, it is replaced by another packet. For this scenario we obtain the following:
\begin{itemize}
 \item  Attempt probability of the AP and the STA
\item  Throughput achieved by the AP and the STA (We will call it saturation throughput, since queues of the AP and the STA are saturated)
\end{itemize}

The paper is organized as follows: Section \ref{sec:related_work}, we discuss related work. In Section \ref{sec:psm_overview} gives the overview of the PSM and the queuing structure at the AP. Section \ref{sec:ieee_overview} gives the overview of the IEEE 802.11 DCF.  In Section \ref{sec:attempt_prob}, we analyze the attempt probabilities of AP and STA. In Section~\ref{sec:sat_throughput}, we obtain the saturation throughput of the AP and the STA. Section~\ref{sec:sim_results} shows the comparison of the analytical and simulation results. Finally, Section~\ref{sec:conclusion} concludes the paper.

\section{Related Work}\label{sec:related_work}
In a seminal paper, Bianchi~\cite{Base:bianchi_2000} proposed an approximate model for the throughput performance of a single-cell IEEE~802.11 network that uses DCF as the medium access mechanism and in which all the nodes are saturated.  Kumar et al.~\cite{Base:kumar_fpa_05} extended the model and provided some new insights. In both of these papers, the authors evaluated the attempt probability $\beta$ (also, the long term attempt rate)  in a system slot as a function of the number of contending nodes. While in this paper, we obtain the attempt probability and saturation throughput for a scenario, which requires  different analysis than presented in ~\cite{Base:bianchi_2000} and~\cite{Base:kumar_fpa_05}, due to the different behavior of the PSM.
In our earlier submission~\cite{link:comsnets}, we focussed on the performance of PSM under application level downlink traffic over TCP. While in this paper, we consider both uplink and downlink traffic, which is different from TCP. We analyze attempt probability of the AP and the STA and saturation throughputs, which makes this paper more basic than our earlier submission~\cite{link:comsnets}. 

 This is the first paper to evaluate this scenario, there has been no attempt in the previous studies to obtain saturation throughput and the attempt probability for the saturated STA in PSM and the saturated AP.
%While, in this paper we consider a single STA in PSM, for this scenario we evaluate the long term attempt rates of the saturated AP and the saturated STA. Due to the PSM protocol, analysis for long term attempt rate of the AP and the STA is different that of proposed in ~\cite{Base:bianchi_2000} and ~\cite{Base:kumar_fpa_05}.

Anastasi et al.~\cite{Analytical:saving_energy_psm_pspoll_anastasi_04}, Lei and Nilsson~\cite{Analytical:bulk_service_lei_nilsson_05}, Baek and Choi~\cite{Analytical:perform_extension_of_nelson_09} and Si et al.~\cite{Analytical:novel_down_link_scheme_pengbo_08} evaluated the energy performance of PSM, but none of them attempt to obtain the saturation attempt rates of the STA and the AP.
Apart from this, in all the above papers, authors consider the PSM protocol implementation which is not practical. 
They consider the following sequence of frame exchanges: First the PSM STA sends the PS-POLL frame through contention, after SIFS AP sends the data packet and after SIFS again the STA sends the MAC ACK. So the AP does not contend to send data. 
In the presence of traffic from the AP to other STAs, when the AP receives the PS-POLL frame, some packets might be already present in the NIC queue of the AP, and these packets need to be sent first. So the above sequence of frame exchange cannot work under the scenario just described. We consider a different implementation of PSM protocol which is explained in the next section.
% We have analyzed the PSM protocol in the presence of traffic from other STAs, which is a more realistic scenario and so we have considered an implementation of the PSM protocol in which the AP contends to send data to PSM STAs.
% Hu et al.~\cite{Analytical:micro_ibss_rong_06} consider STAs in an independent basic service set (IBSS) and evaluate the throughput, delay and the loss rate of the energy characterization as a function of the traffic load, buffer size and other protocol specific parameters. Our work is different from this as we focus on the STAs in an Infrastructure Basic Service Set.

Krashinsky and Balakrishnan~\cite{Experimental:ronny_bsd_05} and Quiao and Shin~\cite{Experimental:smart_psm_quaio_05}  focus on the interaction of the TCP slow start, RTT and PSM.
%They estimates the RTT of current TCP connection and using this information, the sleep wake schedule is made more efficient. 
% Anand et al.~\cite{Experimental:selftuning_night_03} demonstrate the degradation of performance of latency sensitive application like NFS and audio streaming for a STA in PSM. To prevent this degradation, they proposed a self-tuning power management (STPM) algorithm  which accepts inputs from applications.
%  On the basis of the inputs, the algorithm evaluates expected energy and delay incurred in both the modes (CAM and PSM), using which it decides to operate in a particular mode. Our work can complement this by quantifying the exact value of the energy that is consumed while a TCP application is running, so it can help to devise better power management policies. 
Yong et al.~\cite{Experimental:scheduled_psm_yong_07} propose a way to minimize energy and delay by scheduling and informing the  schedule to STAs through beacon frames. 
% The authors show that scheduled PSM improves the performance in terms of energy as it reduces the idle times in the presence of background traffic.
  Tan et al.~\cite{Experimental:psm_throttling_enhua_07} propose to take advantage of throttling done by the TCP server in media streaming applications. In all of these papers, the authors focus on the energy saving either by modifying the PSM protocol~\cite{Experimental:scheduled_psm_yong_07}, or by modifying the sleep wake schedule of the radio depending upon the characteristics of the application level traffic~\cite{Experimental:ronny_bsd_05},~\cite{Experimental:smart_psm_quaio_05} and~\cite{Experimental:psm_throttling_enhua_07}.

\begin{figure}[ht]
\centering
\includegraphics[width= 3 in]{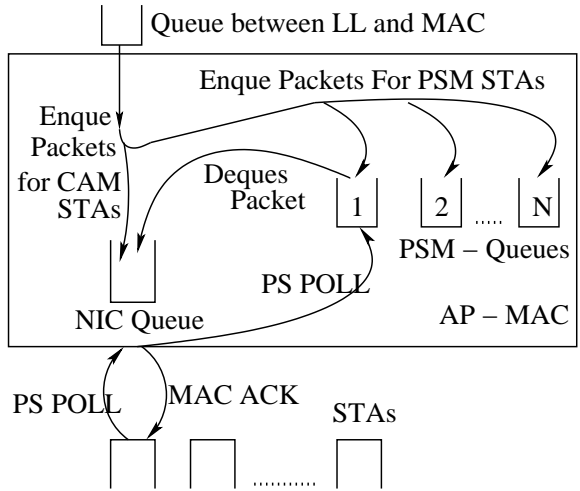} 
\caption{Queuing Structure at AP}
\label{fig:psm_queue}
\end{figure}

\section{PSM - Overview} \label{sec:psm_overview}
In this section, we discuss the implementation of the PSM protocol that we have considered, and the queueing structure at the AP for the packets destined for the PSM clients.
When any STA switches to Power Save Mode (PSM), it goes to sleep state (switches off its radio) and also informs the AP about it. Packets arriving at the AP for PSM STAs are stored in separate queues maintained for each PSM STA; we call them \emph{PSM Queues}, see Fig. \ref{fig:psm_queue}. There is a NIC queue in which MAC PDUs are enqueued for transmission by the PHY layer.
The AP sends beacon frames periodically, through which it informs PSM STAs about the packets enqueued for them. PSM STAs also wake up periodically to listen to the beacon frame. If, on receiving a beacon, an STA sees an indication that there is a packet enqueued for it, then it contends for the medium to send the PS-POLL frame. In reply to the PS-POLL frame, the AP immediately sends a MAC ACK. \emph{This behavior is in contrast to earlier studies where it is assumed that a data packet is sent immediately.} It is not practical to assume that the AP can immediately send a data packet in response to the PS-POLL, since there might be already some packets present in the NIC queue of the AP at the instant when AP receives the PS-POLL frame.

On receiving the PS-POLL from a STA, the AP dequeues the HOL packet from the corresponding PSM queue and enqueues it at the tail of the NIC queue. If the PSM queue of the STA is still non-empty, then the AP sets the More Bit in the dequeued packet to indicate to the STA that there are more packets stored for it at the AP. On receiving this packet, the STA checks the More bit. If it is set then it sends another PS-POLL frame. In this way, the STA does not sleep until its PSM queue at the AP becomes empty. And also, note that each PS-POLL packet permits the STA to download one packet enqueued at the AP. 

Since the PS-POLL is a MAC level packet, it is enqueued at the HOL position in the transmission queue of the STA. If the STA is contending for a packet when a PS-POLL frame is generated, then the STA does not sample a new backoff, but uses the residual backoff of the packet under contention. Further, it is not possible that the STA receives a data packet when it is contending for PS-POLL, because it is only after the PS-POLL is sent, a data packet arrives at the STA. 

There are some situations which are not specified in the protocol but are implementation dependent. Such situations and the assumed behaviors of an STA and the AP are described here.
After sending a PS-POLL, the STA marks its state as \emph{waiting for unicast}. If before the STA receives the unicast packet, the AP transmits a beacon frame and it indicates that there are packets at the AP for this STA, then this STA will not generate another PS-POLL frame. But this may result in a deadlock when the packet that it is meant for is lost, because then the STA will continue to be awake and will not send another PS-POLL. To prevent this situation, a timer is started when the STA sends the PS-POLL, and if the STA does not receive a packet before timer expiry, it goes to the sleep state. Subsequently in the next beacon interval, if the STA gets an indication, then it will send a PS-POLL to retrieve the packet from the AP. Further, if the beacon frames arrives at the STA when it is contending for PS-POLL, then it ignores the beacon frame, because the STA already knows that there is a packet at the AP for it.

\begin{figure*}[ht]
\centering
\includegraphics[width= 7 in]{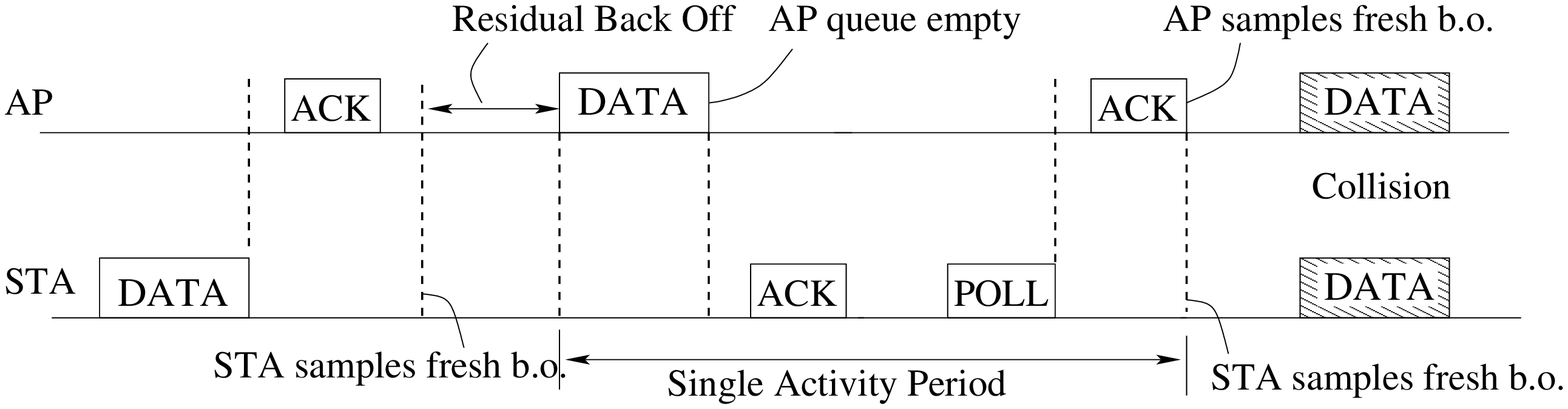} 
\caption{}
\label{fig:s_one}
\end{figure*}

\begin{figure*}[htbp]
\begin{center}
\scalebox{.6}{\input{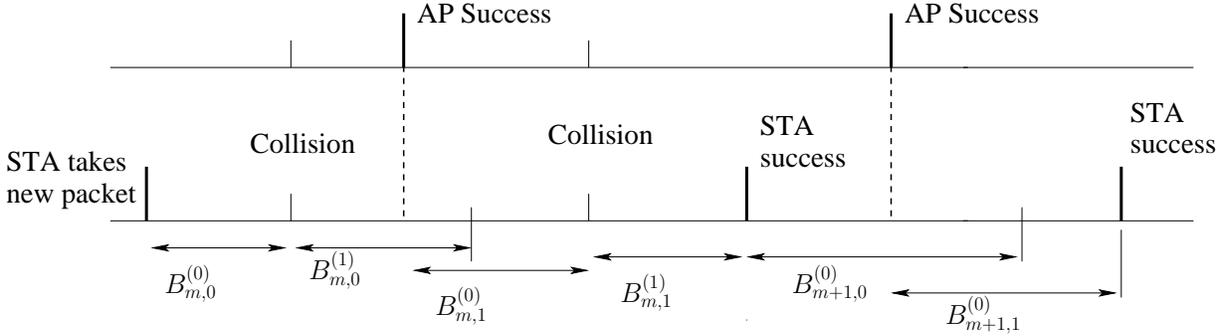}}
 \caption{Showing only the back off duration after removing the activity period}
\label{fig:s_two}
\end{center}
\end{figure*}

\section{IEEE 802.11 overview} \label{sec:ieee_overview} 
In this section we give a brief overview of IEEE 802.11 DCF. We will refer to any wireless entity as node. If any node wants to transmit a packet then it first samples a backoff, which is a uniformly chosen number in between $(0,b_0-1)$, where, $b_0$ is some number defined by the protocol. We call this range as the collision window and the chosen number as backoff counter. In IEEE 802.11, time is divided into slots, and any node can attempt at the boundary of the slot only. After sampling the backoff, node senses the channel, if the channel is idle for more than DIFS than the node starts decrementing the backoff counter. The node decrements the counter by $1$, if it senses the medium idle for a slot. It continues to decrement the backoff counter until it finds that the channel is busy or the counter reaches $0$. On sensing the  medium busy, the node freezes its backoff counter and hereafter continuously senses the medium to become free. After sensing the medium free for DIFS duration the node again starts decrementing the remaining counter. If the counter reaches $0$, than it attempts to transmit the packet. If there is a collision then it increases the collision window and samples a new backoff from this new window. This procedure continues until either the node is able to successfully transmit the packet or it reaches the maximum number of collisions it can experience before the packet is discarded. We denote $K$ the maximum  maximum number of collisions the packet can experience before it is discarded. We denote the initial range of collision window as $(0,b_0-1)$, and after $k$ collisions, collision window as $(0 , b_k-1)$.

\section{Attempt Probability} \label{sec:attempt_prob}
Since the AP is assumed to be saturated, which means it always has a data packet for the STA, so More bit is set in every data packet sent to the STA. On receiving a data packet, the STA has to send a PS-POLL frame. Since the PS-POLL is a MAC level packet, it is enqueued at the HOL position in the transmission queue of the STA. 

To transmit a PS-POLL, as discussed earlier also, the STA does not sample a new backoff, but uses the residual backoff of the packet for which it was contending at the instant when the STA received a data packet from the AP (see Fig.~\ref{fig:s_one}). Although the AP is saturated, but all the packets lies in the PSM queue of the STA at the AP, and it is only after it receives a PS-POLL, a packet arrives in the NIC queue of the AP. So while the STA is counting down the backoff counter for the PS-POLL, the NIC queue of the AP is empty, so the STA transmits a PS-POLL without any possibility of collision. We treat the transmission of a data packet by the AP and the subsequent PS-POLL frame by the STA as a single activity period, this activity period also includes the back off counted for the PS-POLL frames, as shown in the Fig~\ref{fig:s_one}.

After the STA transmits a PS-POLL, it samples a fresh back-off from the initial collision window (0, $b_0 -1 $) and again starts contending for the data packet. Thus we see that when the AP successfully transmits a data packet, it results in the STA sampling a fresh backoff for its packet. As a result the STA has to countdown more backoff than the AP to transmit a data packet, due to which the attempt probability of the STA is less than that of the AP. Since the STA uses the residual backoff to transmit a PS-POLL and there is no contention to transmit it, because of this the attempt rate to transmit a PS-POLL is higher than than that of the STA and the AP to transmit a data packet. In this section, we evaluate long term attempt rates of the AP to transmit data packet and the attempt rate of the STA to transmit data packet and PS-POLL. In next section, we will evaluate the saturation throughput of the AP and the STA.

We consider the actvity duration as the trasnmission durations of all the frames, collisons, SIFS, DIFS, EIFS. we remove all the activity durations from the figure~\ref{fig:s_one}, and leaving behind only the backoff time as shown in Fig.~\ref{fig:s_two}. Note that it is the restricted back off time, since we consider the residual backoff counted to transmit the PS-POLL in the activity period as shown in the Fig~\ref{fig:s_one}.
\\
Define the following, for $0 \leq k \leq K$; $m \geq 0$, $j \geq 0$ \\
$ B_{m,j}^{(k)}$ := It is the back off sampled by the STA, after the $j^{th}$ AP interruption during the contention of its $m^{th}$ data packet and the $k^{th}$ back off duaration.
\\
Consider Fig.~\ref{fig:s_two}, the STA takes a new data packet ($m^{th}$) and samples a backoff $B_{m,0}^{(0)}$. After decrementing this backoff there is a collision, so the STA samples a new back off $B_{m,0}^{(1)}$. When the STA is counting down for $B_{m,0}^{(1)}$, there is a AP success and back off of the STA is interrupted, as the remaining backoff of $B_{m,0}^{(1)}$ is used by the STA to transmit PS-POLL. The STA now samples a new back off $B_{m,1}^{(0)}$ and this goes on.

Assume the AP attempts in this restricted back off times at rate $\beta_{a}$ per slot and the STA attempts with the rate $\beta_{s}$ per slot.

Let the distribution of the $k^{th}$ back off duration is:
\begin{equation}
 p_j^{(k)} = \frac{1}{b_k}, \,\,\mbox{for}\,\, 0 \leq j \leq b_k - 1, 0 \leq k \leq K
\end{equation}

Lets denote $b^{(k)}$ the expectation of the $k^{th}$ back off duration, which is given by:
\begin{equation}
 b^{(k)} = \sum_{j = 0}^{b_k-1}j\frac{1}{b_k} = \frac{b_k -1}{2}
\end{equation}

Assume the follwoing 
\begin{itemize}
 \item The nodes attempt as a bernoulli process during the restricted back off times
\item The attempt process is independent
\end{itemize}
Let $n$ be the index of restricted back off times, we see that

\begin{align}
 \beta_{a} &= \lim_{n \to \infty}\frac{A_{a}(n)}{n} \\
\beta_{s} &= \lim_{n \to \infty}\frac{A_{s}(n)}{n} 
\end{align}
Where, $A_{a}(n)$ and $A_{s}(n)$ are the attempts count untill $n$, in the restricted back off times.
The success instants of the AP (in restricted back off times) are renewal instants. 

Denote $A_{a,m}$, $A_{s,m}$ the number of attempts made by the AP and the STA respectively to transmit their $m^{th}$ data packet. Lets denote  $X_{a,m}$, $X_{s,m}$ the number of back off slots decremented by the AP and the STA respectively to trasnmit their  $m^{th}$ packet. Note that $X_{s,m}$ does not include the back off used to trasnmit PS-POLL frames.

Using the renewal reward theorm~\cite{book:book_rw_wolff}, following can be written for $\beta_{a}$ and $\beta_{s}$:

\begin{align}
\begin{split}
 \beta_{a} &= \lim_{n \to \infty}\frac{A_{a}(n)}{n}  = \frac{E[A_{a,1}]}{E[X_{a,1}]} \\
\beta_{s} &= \lim_{n \to \infty}\frac{A_{s}(n)}{B_{s}(n)} 
	    = \frac{E[A_{s,1}]}{E[X_{s,1}]}
\end{split}
\end{align}

\subsection{Attempt Probability of the AP}
Since the STA is attempting with probability $\beta_{s}$ in every slot.
So, given the AP attempts in any slot the probability that there will be a collision is $\beta_{s}$. Recalling, $K$ is the maximum number of collisions a packet can experience before it is discarded. The expected number of attempts between any two success of the AP is given by:

\begin{align}
 E[A_{a,1}] = 1 + \beta_{s} + \beta_{s}^2 + ... + \beta_{s}^{K}
\end{align}

Expected backoff decremented by the AP between two success of it, can be given by the following equation:

\begin{align}
 E[X_{a,1}] &= b^{(0)} + b^{(1)}\beta_{s} + b^{(2)}\beta_{s}^2 + \ldots + b^{(K)}\beta_{s}^{K}
\end{align}
Detailed derivation of the above expressions can be found in~\cite{Base:kumar_fpa_05}.
\subsection{Attempt Probability of the STA}
Lets denote the $\tilde{b}^{(k)}$ as the mean duration of the $k^{th}$ back off duration to transmit any packet, taking into account of the AP interruption and the back off restart. So the $E[X_{s,1}]$ = $\tilde{b}^{(0)}$.

In each slot the AP attempts with probability $\beta_{a}$, and given the STA attempts in any slot, the probability that there is a collision is $\beta_{s}$. 

\begin{figure}
\centering
\includegraphics[width= 2.5 in]{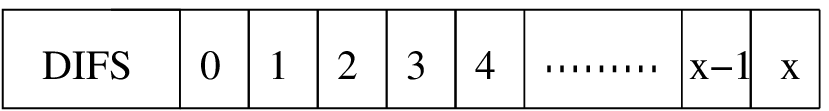} 
\caption{}
\label{fig:slot_mark}
\end{figure}

We start from the point, where the STA has meet $k^{th}$ collsion while transmitting a packet. After this point, the STA waits for DIFS duration and samples a new backoff from the interval $(0,b_k-1)$. As shown in the fig.~\ref{fig:slot_mark}, we have numbered the slots after DIFS duration as $0, 1, 2, 3 ...$.

Lets assume that the STA samples the back off $x$, which means the STA will attempt to transmit in slot number $x$. There are three possible cases:

\begin{itemize}
\item AP success -- In this case the AP attempts in any slot $i$ $\in$ $(0,x-1)$, probability of this event is $(1 - \beta_{a})^{i} \beta_{a}$ and the number of slots counted by the STA is $i$. Since the STA restarts the back off, so from here onwards the expected back off duration is $\tilde{b}^{(0)}$. 
\item STA sucess -- In this case the AP does not attempt in any slot $i$ $\in$ $(0,x)$, probability of this event is $(1 - \beta_{a})^{x+1}$ and the number of slots counted by the STA is $x$. 
\item Collision -- In this case the AP sattempts in slot $x$, probability of this event is $(1 - \beta_{a})^{x} \beta_{a}$. The number of slots counted by the STA is $x$, and for $k<K$ the expected back off duration from here onwards is  $\tilde{b}^{(k+1)}$ and for $k=K$ it is $0$.
\end{itemize}

Expression for $\tilde{b}^{(k)}$ can be written as follows:

\begin{align}
\begin{split}
\tilde{b}^{(k)} &= \sum_{x=0}^{b_k - 1} p_x^{(k)} \left[\sum_{i = 0}^{x-1} (1 - \beta_{a})^{i}\beta_{a}(i + \tilde{b}^{(0)})\right] \\& + \sum_{x=0}^{b_k - 1} p_x^{(k)} (1 - \beta_{a})^{x}\beta_{a}(x + I_{\{k < K\}}\tilde{b}^{(k+1)}) \\& +
\sum_{x=1}^{b_k - 1} p_x^{(k)} (1 - \beta_{a})^{x+1}(x)
\end{split}
\end{align}

Let us denote the following:
\begin{align}
\begin{split}
X_k &=  1 - \frac{1 - (1 - \beta_{a})^{b_k}}{b_k\beta_{a}} \\
Y_k &=  \frac{1 - (1 - \beta_{a})^{b_k-1}}{b_k}\\
Z_k &= \sum_{x=0}^{b_k-1}\left[\frac{1}{b_k}\sum_{i=0}^{x-1}(1 - \beta_{a})^{i}\beta_{a}(i) \right]\\ &+  \sum_{x=0}^{b_k-1}\frac{1}{b_k}\left[ (1 -  \beta_{a})^{x}\beta_{a}(x) + (1 - \beta_{a})^{x+1}(x) \right]
\end{split}
\end{align}

Using the above notations, $\tilde{b}^{(k)}$ can be written as follows,
\begin{align}
\begin{split}
 \tilde{b}^{(k)} &= X_k\tilde{b}^{(0)}+ Y_k\tilde{b}^{(k+1)} + Z_k \qquad \mbox{for $k < K$} \\
&= X_K\tilde{b}^{(0)}+  Z_K \qquad \qquad  \qquad \mbox{for $k = K$} \\&
\end{split}
\end{align}

Using the above recursive equation, following expression can be written for $\tilde{b}^{(0)}$:
\begin{align}
\begin{split}
 \tilde{b}^{(0)} &= \sum_{k=0}^{K} X_k\prod_{l=0}^{k-1}Y_l \tilde{b}^{(0)} + \sum_{k=0}^{K} Z_k\prod_{l=0}^{k-1}Y_l 
\\&=\frac{\sum_{k=0}^{K} Z_k\prod_{l=0}^{k-1}Y_l}{ 1- \sum_{k=0}^{K} X_k\prod_{l=0}^{k-1}Y_l}
\end{split}
\end{align}

% $E[B_{STA}] = E_0[B]$
Expected number of attempts between two success of the STA is given by the following equation:

\begin{align}
 E[A_{s,1}] = 1 + \beta_{a} + \beta_{a}^2 + ... + \beta_{a}^{K}
\end{align}
Detailed derivation of the above expression can be found in~\cite{Base:kumar_fpa_05}.

\subsection{Attempt probability to transmit PS-POLL frame}
% Since the STA does not contend to send PS-POLL and it uses the residual backoff of the packet under contention when it received the data packet from the AP. Hence attempt rate of the STA to transmit a PS-POLL is higher than that of the AP and the STA to transmit  data packet. In this section, we evaluate this attempt rate.

Lets denote $X_{ps,m}$ the total backoff decremented by the STA for transmitting PS-POLLs, when it was trying to send its $m^{th}$ packet. Lets denote $A_{ps,m}$ the total number of PS-POLLs sent by the STA during the period when it was trying to send its $m^{th}$ packet. We are interested in finding out the attempt rate to send PS-POLLs, which can be written as follows:
\begin{align}
\begin{split}
 \beta_{ps} &= \frac{\sum_{m=0}^{\infty}A_{ps,m}}{\sum_{m=0}^{\infty}X_{ps,m}}
             = \frac{\lim_{r \to \infty}\frac{1}{r}\sum_{m=0}^{r}A_{ps,m}}{\lim_{r \to \infty}\frac{1}{r}\sum_{m=0}^{r}X_{ps,m}}
 	     \\&= \frac{E[A_{ps,1}]}{E[X_{ps,1}]}
\end{split}
\end{align}

\subsubsection{Calculation of $E[X_{ps,1}]$}
Denote $\bps$ the mean backoff duration spent in transmitting PS-POLLs, after the STA has suffered $k$ collisions of a data packet. So $E[X_{ps,1}]$ equal to $\tilde{b}_{ps}^{(0)}$.

By considering the cases, as done in the previous section following gives $\bps$ :
\begin{align}
\begin{split}
\bps &= \sum_{x=0}^{b_k - 1} p_x^{(k)} \left[\sum_{i = 0}^{x-1} (1 - \beta_{a})^{i}\beta_{a}((x-i) + \tilde{b}_{ps}^{(0)})\right] \\& + \sum_{x=0}^{b_k - 1} p_x^{(k)} (1 - \beta_{a})^{x}\beta_{a}(0 + I_{\{k < K\}}\tilde{b}_{ps}^{(k+1)}) \\& +
\sum_{x=0}^{b_k - 1} p_x^{(k)} (1 - \beta_{a})^{x+1}(0)
\end{split}
\end{align}

In the similar way, following gives the expression of $\tilde{b}_{ps}^{(0)}$:
\begin{align}
\begin{split}
 \tilde{b}_{ps}^{(0)}&= \sum_{k=0}^{K} X_k\prod_{l=0}^{k-1}Y_l \tilde{b}_{ps}^{(0)} + \sum_{k=0}^{K} Z_k^{(1)}\prod_{l=0}^{k-1}Y_l 
\\&=\frac{\sum_{k=0}^{K} Z_k^{(1)}\prod_{l=0}^{k-1}Y_l}{ 1- \sum_{k=0}^{K} X_k\prod_{l=0}^{k-1}Y_l}
\end{split}
\end{align}

where,
\begin{align}
\begin{split}
Z_k^{(1)} &= \sum_{x=0}^{b_k-1}\left[\frac{1}{b_k}\sum_{i=0}^{x-1}(1 - \beta_{a})^{i}\beta_{a}(x-i)\right]
\end{split}
\end{align}

\subsubsection{Calculation of $E[A_{ps}]$}
Denote $\aps$ the number of successful transmission of the AP in between two successful transmission of the STA, after the STA has suffered $k$ collisions of a data packet. 
By considering the cases, as done in the previous section following equation gives the expected number of data transmissions by the AP:
Expression for $\aps$ can be written as follows:
\begin{align}
\begin{split}
\aps &= \sum_{x=0}^{b_k - 1} p_x^{(k)} \left[\sum_{i = 0}^{x-1} (1 - \beta_{a})^{i}\beta_{a}(1 + \tilde{a}_{ps}^{(0)})\right] \\& + \sum_{x=0}^{b_k - 1} p_x^{(k)} (1 - \beta_{a})^{x}\beta_{a}(0 + I_{\{k < K\}}\tilde{a}_{ps}^{(k+1)}) \\& +
\sum_{x=0}^{b_k - 1} p_x^{(k)} (1 - \beta_{a})^{x+1}(0)
\end{split}
\end{align}
Following expression gives the value of $\tilde{a}_{ps}^{(0)}$:
\begin{align}
\begin{split}
 \tilde{a}_{ps}^{(0)} &= \sum_{k=0}^{K} X_k\prod_{l=0}^{k-1}Y_l \tilde{a}_{ps}^{(0)} + \sum_{k=0}^{K} Z_k^{(2)}\prod_{l=0}^{k-1}Y_l 
\\&=\frac{\sum_{k=0}^{K} Z_k^{(2)}\prod_{l=0}^{k-1}Y_l}{ 1- \sum_{k=0}^{K} X_k\prod_{l=0}^{k-1}Y_l}
\end{split}
\end{align}
where, $Z_K^{(2)}$ is given by the following equation:
\begin{align}
\begin{split}
Z_k^{(2)} &= \sum_{x=0}^{b_k-1}\left[\frac{1}{b_k}\sum_{i=0}^{x-1}(1 - \beta_{a})^{i}\beta_{a}(1)\right]
\end{split}
\end{align}

\begin{figure*}
\subfloat[Attempt Probability of the AP]{\label{fig:att_ap}\includegraphics[width=3.5 in]{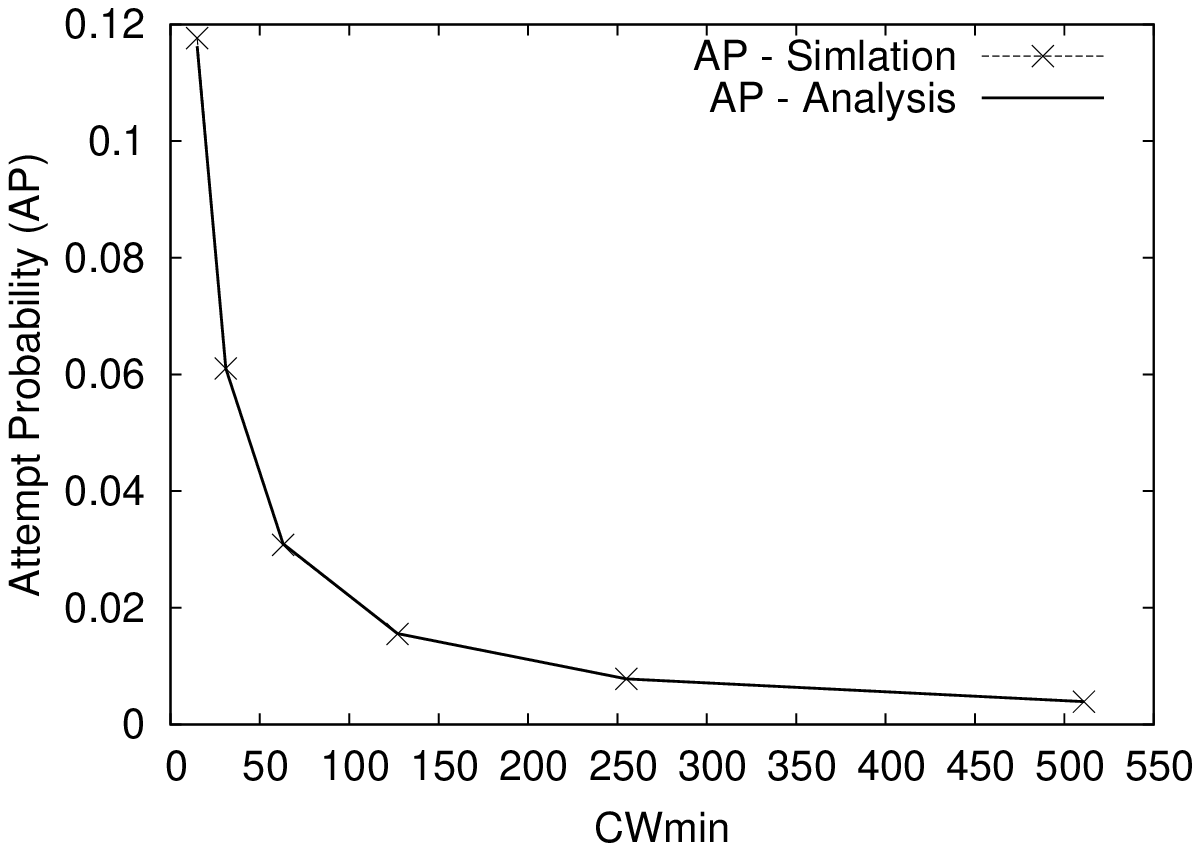}}
 \subfloat[Attempt probability of the STA to transmit data packet]{\label{fig:att_sta_data}\includegraphics[width=3.5 in]{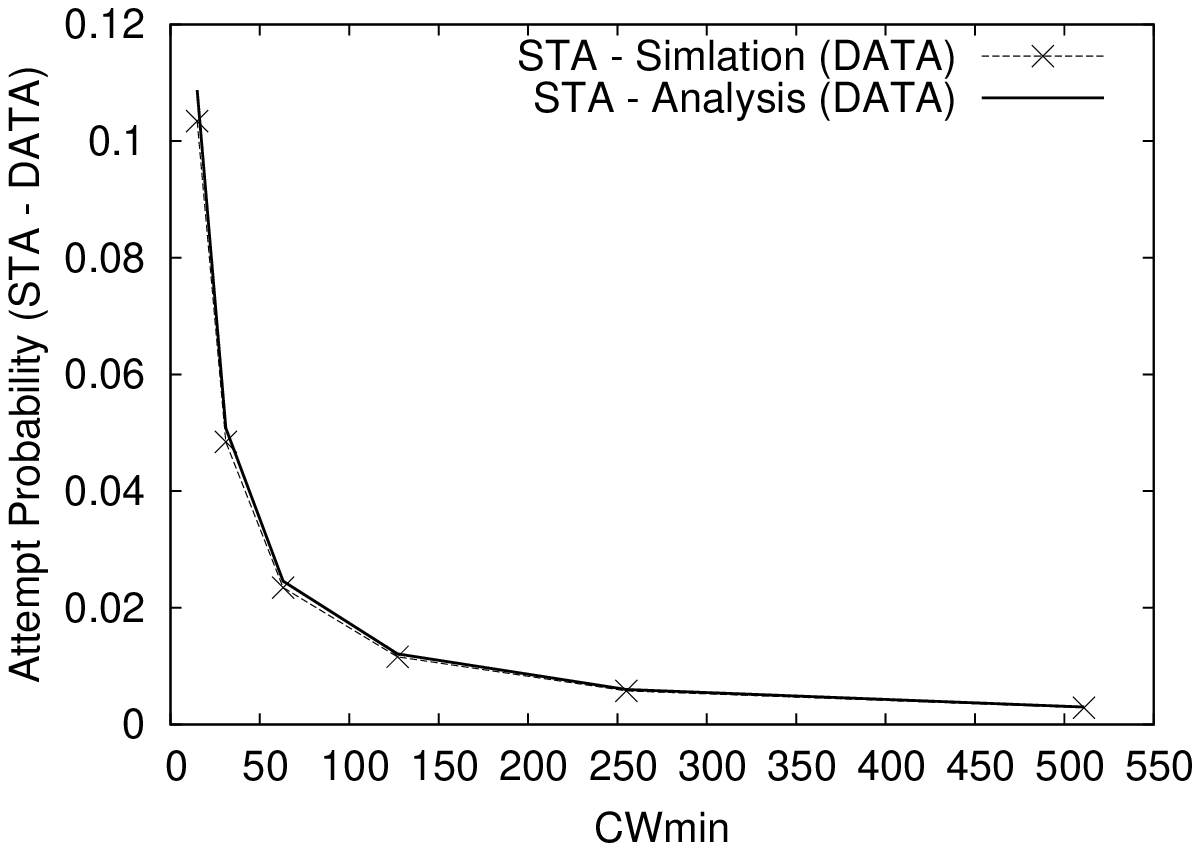}}
\caption{}
\end{figure*}

\begin{figure*}
\subfloat[Attempt Probability of the STA to transmit PS-POLL frame]{\label{fig:att_sta_pspl}\includegraphics[width=3.5 in]{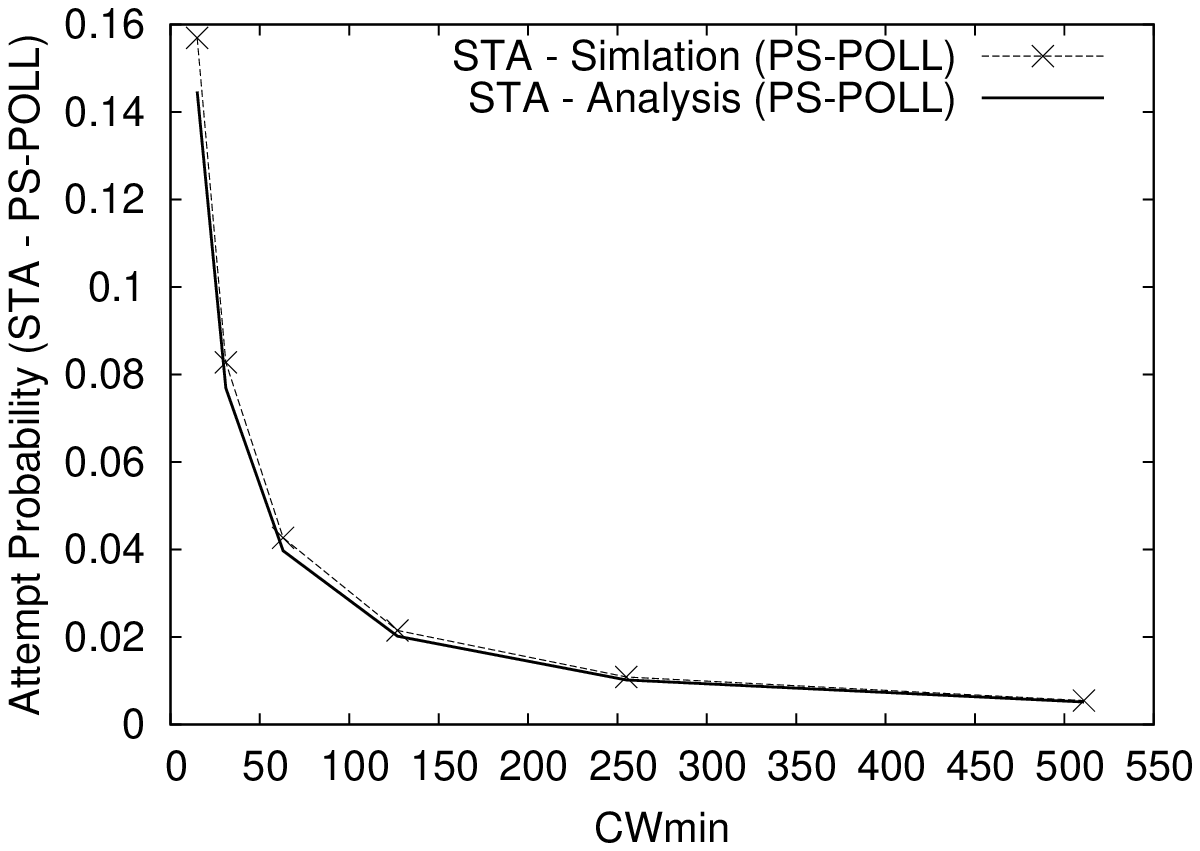}}
\subfloat[Collision Probability]{\label{fig:att_prob_coll}\includegraphics[width=3.5 in]{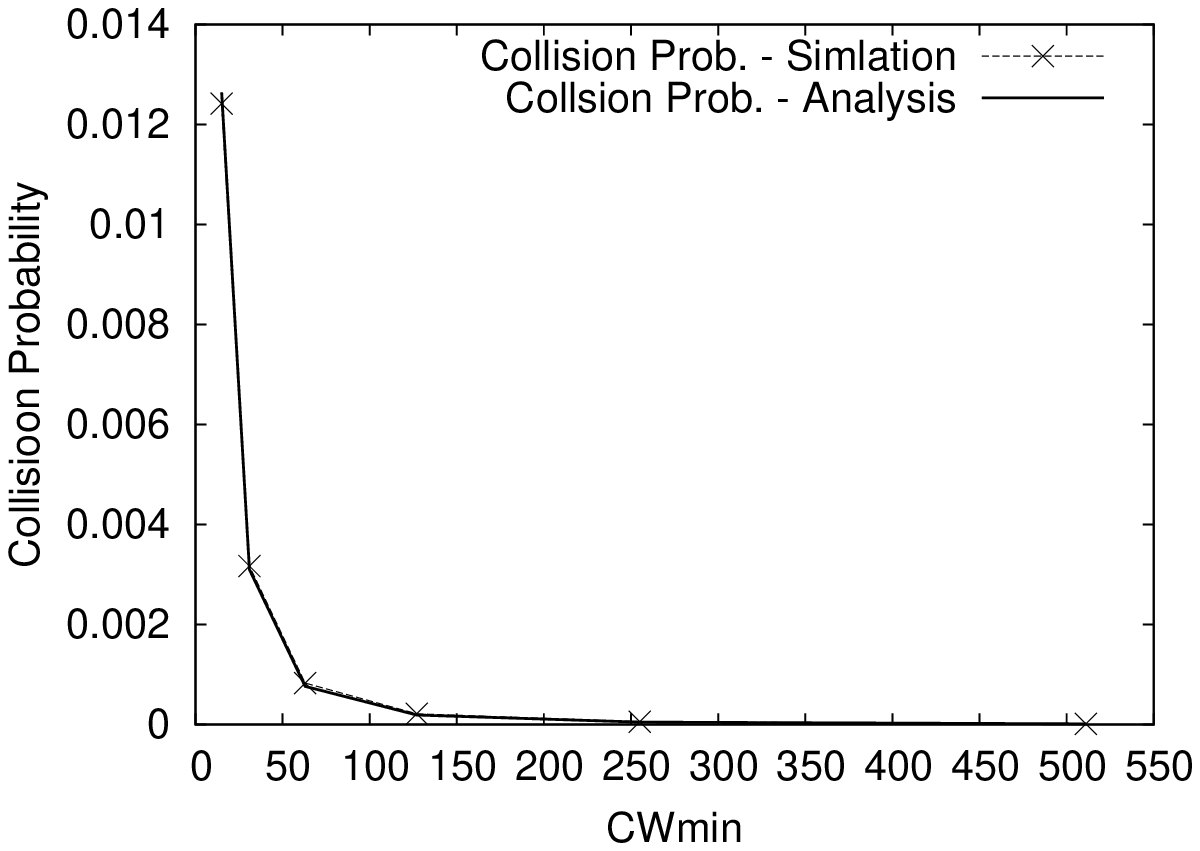}}
\caption{}
\end{figure*}

\section{Saturation Throughput} \label{sec:sat_throughput}
In this section, we analyze the saturation throughput obtained by the AP and the STA. Denote $N_{ap}(t)$ and $N_{sta}(t)$ the number of successful transmissions by the AP and the STA, in the time interval $(0,t)$. Following equations give the throughput achieved by the STA and the AP:

\begin{align}
\begin{split}
 \Theta_{AP} &= \lim_{t \to \infty}\frac{N_{AP}(t)}{t} \\
\Theta_{STA} &= \lim_{t \to \infty}\frac{N_{STA}(t)}{t} 
\end{split}
\end{align}

Consider the following instants:
1) End of successful transmissions of the data packet either by the AP or the STA,
2) End of idle slots,
3) End of collisions.
Let us call the $l^{th}$ such instant as $M_l$. These points are renewal points, for the process $N_{ap}(t)$ and $N_{sta}(t)$. Define $T_l = M_{l} - M_{l-1}$. For any $(M_{l-1},M_{l})$, define $H_{AP,l}$, it is $1$ if there is a successful transmission of data packet by the AP else $0$. Define $H_{STA,l}$ it is $1$ if if there is a successful transmission by the the STA in $(M_{l-1},M_{l})$ else $0$. By applying Renewal reward theorem~\cite{book:book_rw_wolff}, following equations gives the saturation throughput of the STA and the AP:

\begin{align}
\begin{split}
 &\Theta_{AP} = \frac{E[H_{AP}]}{E[T]} \\
& = \frac{P_{sAP}}{P_{sAP}(T_{sAP} ) + P_{sSTA}T_{sSTA} + P_{idle}\delta + P_cT_c }\\
&\Theta_{STA} = \frac{E[H_{STA}]}{E[T]} \\
& = \frac{P_{sSTA}}{P_{sAP}(T_{sAP} ) + P_{sSTA}T_{sSTA} + P_{idle}\delta + P_cT_c }
\end{split}
\end{align}

Since, a transmission of the AP is accompanied by the transmission of a PS-POLL. As discussed earlier, the STA does not have to contend to transmit a PS-POLL, but the STA still has to wait till the residual backoff of the data packet becomes $0$. To account for this residual backoff, we evaluated the $\beta_{ps}$, the expected time taken to transmit PS-POLL is given by:
\begin{align}
\begin{split}
E[T_{PSPL}] &=  (1 - \beta_{pspl})(\delta + E[T_{PSPL}]) + \beta_{pspl}T_{sPSPL}  \\
&= T_{sPSPL}  + \frac{(1 - \beta_{pspl})}{\beta_{pspl}}\delta 
\end{split}
\end{align}
The above equations uses the following notations, and these uses the 802.11 parameters shown in the table~\ref{tab:param}.
\begin{tabular}{ p{1.1 cm}  p{6.5cm}}
$P_{sAP}$  & It is the total probability that AP wins the contention = $\beta_{a}(1 - \beta_{s})$\\
$P_{sSTA}$ & It is the total probability that STA wins the contention = $\beta_{s}(1 - \beta_{a})$\\
$P_{c}$    & It is the total probability that there is a collision = $\beta_{s}\beta_{a}$\\
$P_{idle}$ & It is the total probability that the channel is idle = $(1 - \beta_{s})(1 - \beta_{a})$\\
\end{tabular}

\begin{tabular}{ p{1.1 cm}  p{6.5cm}}
$T_{sAP}$ & It is the time spent when the AP wins contention = $SIFS + DIFS + T_{ACK} +  T_{AP,d} + E[T_{PSPL}]$\\
$T_{sSTA}$  & It is the time spent when the STA wins contention= $SIFS + DIFS + T_{ACK} + T_{STA,d}$\\
$T_{c}$   &It is the time spent when there is a collision= $max(T_{AP,d},T_{STA,d}) + EIFS$\\
\end{tabular}

\begin{align*}
T_{sPSPL}& = T_{PSPL} + SIFS + DIFS + T_{ACK} \\
T_{ACK}  & = T_P + T_{PHY} + \frac{L_{ACK}}{C_c}\\
T_{PSPL} & = T_P + T_{PHY} +  \frac{L_{PSPL}}{C_c}\\
T_{AP,d} & = T_P + T_{PHY} +  \frac{L_{MAC}+L_{AP}}{C_d}\\
T_{STA,d} & = T_P + T_{PHY} +  \frac{L_{MAC}+L_{STA}}{C_d}\\
\end{align*}

\begin{figure}[h]
   \centering
  \includegraphics[width = 3.5 in]{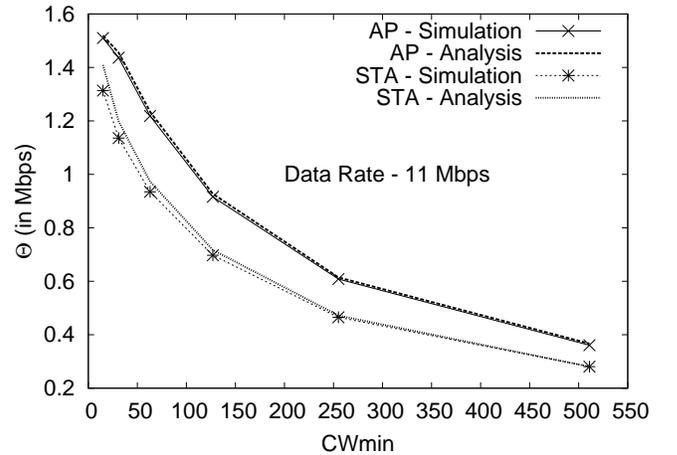}
 \caption{Throughput -- AP and STA (11 Mbps)}
   \label{fig:throughput_ap_sta}
 \end{figure}

\section{Simulation Results} \label{sec:sim_results}
Simulation results are obtained using NS-2.33~\cite{simulator:ns_2} and the various parameters used are taken from the 802.11b standard (given in Table ~\ref{tab:param}). Data packet size transmitted by the AP and the STA is taken as 512 bytes. RTS threshold taken as 600 bytes, which means that the AP and the STA sends the data packet by using basic scheme. Maximum number of attempts ($K$) that the STA or AP can make to transmit a packet is taken as $7$. $b_k = 2^{5+k}$ for $0\leq k \leq5$ and for $5<k\leq7$ $b_k = 2^{10}$.
To verify our analysis, we plotted the attempt probability of the STA and AP to transmit the data packet for varying initial range of collision window $(0,b_0-1)$, in the figures it is written as CWmin (minimum collision window). Figure~\ref{fig:att_sta_data} and~\ref{fig:att_ap} shows the attempt probability of the STA and the AP to transmit data packet. we can see that the long term attempt rate of the STA is more than that of the STA, as a result of which the throughput of the AP is higher than that of the STA as shown in the Figure~\ref{fig:throughput_ap_sta}.
Figure~\ref{fig:att_sta_pspl} shows the long term attempt rate of the STA to transmit PS-POLLs, on comparing it with figures~\ref{fig:att_sta_data} and~\ref{fig:att_ap}, it is clear that the attempt rate of the STA to transmit PS-POLL is higher than that of the the AP and the STA. Figure~ \ref{fig:att_prob_coll} shows the collision probability which is the product of $\beta_{a}$ and $\beta_{s}$ as a function of the minimum collision window ($CWmin = b_0$). We can see that there is a close match between the analytical values and the simulation results which shows the correctness of our analysis.

\section{Conclusion} \label{sec:conclusion}
In this paper, we have presented the analytical modeling of the saturation throughput, when a single STA is associated to the AP, and the STA is in PSM. Due to the different behavior of the STA in PSM from STA in CAM, different analysis is required. We have analyzed the long term attempt rates of the STA and the AP and using them we obtained the saturation throughputs. We have shown that our modeling is correct as the analytical results matches well with the simulation results. As far as the previous literature is concerned, it is the first work which has modeled the scenario presented in the paper. And also, we have considered a more realistic implementation of the PSM, which is different from the PSM protocol considered in previous papers. Our future work will focus on modeling the scenarios, when there are more than one STA in PSM. We will also analyze the mixed scenario in which there are some STAs in CAM and some in PSM.

\begin{table}
\caption{Parameters}
\begin{center}
\renewcommand{\arraystretch}{1.3}
\begin{tabular}{|p{3 cm}|l|l|}
\hline
\multicolumn{3}{|c|}{Parameter used} \\ \hline
Parameter & Symbol & Value\\ \hline
PHY data rate & $C_d$ & 11 Mbps\\
Control rate & $C_c$ & 2 Mbps\\
PLCP Header time & $T_P$ & 144$\mu$s \\
PHY Header time & $T_{PHY }$ & 48$\mu$s\\
MAC Header Size & $L_{MAC}$ & 34 bytes \\
% RTS Packet Size & $L_{RTS}$ & 20 bytes \\
PS-POLL Packet Size & $L_{PS-POLL}$ & 20 bytes \\
% CTS Packet Size & $L_{CTS}$ & 14 bytes \\
MAC ACK Header Size & $L_{ACK}$ & 14 bytes \\
% IP Header & $L_{IPH}$ & 20 bytes \\
% TCP Header & $L_{TCPH}$ & 20 bytes \\
% TCP ACK packet size & $L_{TCP-ACK}$ & 20 bytes\\
Data packet size transmitted by AP & $L_{AP}$ & 512 bytes\\
Data packet size transmitted by STA & $L_{STA}$ & 512 bytes\\
System Slot time & $\delta$ & 20$\mu$s \\
DIFS Time & $T_{DIFS}$ & 50$\mu$s  \\
SIFS Time & $T_{SIFS}$ & 10$\mu$s  \\
EIFS Time & $T_{EIFS}$ & 364$\mu$s  \\
% \hline
% Idle Current & $J_{Id}$ & 200 mA \\
% Receive \& Decode Current & $J_{RxD}$ & 200 mA \\
% Listen Current & $J_{Ls}$ & 200 mA \\
% Transmit Current & $J_{Tx}$ & 300 mA\\
\hline
\end{tabular} \label{tab:param}
\end{center}
\end{table}

 %\bibliographystyle{./IEEEtranBST1/IEEEtran}
 %\bibliography{./IEEEtranBST1/IEEEabrv,./IEEEtranBST1/IEEEexample}
%\include{agrawal_etal_09_wisard.bbl}

\appendix
\subsubsection{Calculation of $E[X_{ps,1}]$}
% In this section, we will evaluate the expected backoff counter decremented by the STA to transmit PS-POLLs in between two successes of the STA.
Denote $\bps$ the mean backoff duration spent in transmitting PS-POLLs, after the STA has suffered $k$ collisions of a data packet. 
Then $\tilde{b}_{ps}^{(0)}$ is equal to $E[X_{ps,1}]$.

% If the backoff sampled by the STA is $x$, and if the AP attempts in the slot $i\,\,(<\,x)$ 
% %with probability $(1 - \beta_{a})^{k-1}\beta_{a}$, 
% then the residual backoff used to transmit PS-POLL is $(x-i)$.
%   By considering the cases as done in the previous section following equation gives the expected backoff decremented by the STA to transmit PS-POLLs:
Lets assume that the STA samples the back off $x$, in that case there are three possible cases:

\begin{itemize}
\item AP success -- In this case the AP should attempt in any slot $i$ $\in$ $(0,x-1)$. Probability that the AP attmepts in $i^{th}$ slot is  $(1 - \beta_{a})^{i} \beta_{a}$ and the number of slots counted by the STA to transmit PS-POLL is $x-i$. Since the STA will restart the back off so from here onwards the expected back off duration is $\bps$. 
\item STA sucess -- In this case the AP should not attempt in any slot $i$ $\in$ $(0,x)$. Probability of this event is $(1 - \beta_{a})^{x+1}$ and the number of slots counted by the STA to transmit PS-POLL is $0$. 
\item Collision -- In this case the AP should attempt in slot $x$. Probability of this event is $(1 - \beta_{a})^{x} \beta_{a}$ and the number of slots counted by the STA is $0$, and for $k<K$ the expected back off duration from here onwards is  $\tilde{b}_{ps}^{(k+1)}$ and for $k=K$ is $0$.
\end{itemize}

So the expression for $\bps$ can be written as follows:
\begin{align}
\begin{split}
\bps &= \sum_{x=0}^{b_k - 1} p_x^{(k)} \left[\sum_{i = 0}^{x-1} (1 - \beta_{a})^{i}\beta_{a}((x-i) + \tilde{b}_{ps}^{(0)})\right] \\& + \sum_{x=0}^{b_k - 1} p_x^{(k)} (1 - \beta_{a})^{x}\beta_{a}(0 + I_{\{k < K\}}\tilde{b}_{ps}^{(k+1)}) \\& +
\sum_{x=0}^{b_k - 1} p_x^{(k)} (1 - \beta_{a})^{x+1}(0)
\end{split}
\end{align}

\begin{align}
\begin{split}
\bps &= X_k\tilde{b}_{ps}^{(0)}+ Y_k\tilde{b}_{ps}^{(k+1)} + Z_k^{(1)} \qquad \mbox{for $k < K$} \\
&= X_K\tilde{b}_{ps}^{(0)}+  Z_K^{(1)} \qquad \qquad \qquad  \qquad \mbox{for $k = K$} \\&
\end{split}
\end{align}

where,

\begin{align}
\begin{split}
Z_k^{(1)} &= \sum_{x=0}^{b_k-1}\left[\frac{1}{b_k}\sum_{i=0}^{x-1}(1 - \beta_{a})^{i}\beta_{a}(x-i)\right]
\end{split}
\end{align}

Using the above recursive equation, following equation can be written for $\tilde{b}_{ps}^{(0)}$:
\begin{align}
\begin{split}
 \tilde{b}_{ps}^{(0)}&= \sum_{k=0}^{K} X_k\prod_{l=0}^{k-1}Y_l \tilde{b}_{ps}^{(0)} + \sum_{k=0}^{K} Z_k^{(1)}\prod_{l=0}^{k-1}Y_l 
\\&=\frac{\sum_{k=0}^{K} Z_k^{(1)}\prod_{l=0}^{k-1}Y_l}{ 1- \sum_{k=0}^{K} X_k\prod_{l=0}^{k-1}Y_l}
\end{split}
\end{align}

\subsubsection{Calculation of $E[A_{ps}]$}
Denote $\aps$ the number of successful transmission of the AP in between two successful transmission of the STA, after the STA has suffered $k$ collisions of a data packet. Then $E[A_{ps}]$ is equal to $\tilde{b}_{ps}^{(0)}$.

Lets assume that the STA samples the back off $x$, in that case there are three possible cases:
\begin{itemize}
\item AP success -- In this case the AP should attempt in any slot $i$ $\in$ $(0,x-1)$. Probability that the AP attmepts in $i^{th}$ slot is  $(1 - \beta_{a})^{i} \beta_{a}$ and the number of AP success is $1$. Since the STA will restart the back off so from here onwards the expected number of AP success is  $\aps$.
\item STA sucess -- In this case the AP should not attempt in any slot $i$ $\in$ $(0,x)$. Probability of this event is $(1 - \beta_{a})^{x+1}$ and the number of AP success is  $0$.
\item Collision -- In this case the AP should attempt in slot $x$. Probability of this event is $(1 - \beta_{a})^{x} \beta_{a}$ and $k<K$ the expected back off duration from here onwards is  $\tilde{a}_{ps}^{(k+1)}$ and for $k=K$ is $0$.
\end{itemize}

So the expression for $\aps$ can be written as follows:

\begin{align}
\begin{split}
\aps &= \sum_{x=0}^{b_k - 1} p_x^{(k)} \left[\sum_{i = 0}^{x-1} (1 - \beta_{a})^{i}\beta_{a}((x-i) + \tilde{a}_{ps}^{(0)})\right] \\& + \sum_{x=0}^{b_k - 1} p_x^{(k)} (1 - \beta_{a})^{x}\beta_{a}(0 + I_{\{k < K\}}\tilde{a}_{ps}^{(k+1)}) \\& +
\sum_{x=0}^{b_k - 1} p_x^{(k)} (1 - \beta_{a})^{x+1}(0)
\end{split}
\end{align}

\begin{align}
\begin{split}
\aps &= X_k\tilde{a}_{ps}^{(0)}+ Y_k\tilde{a}_{ps}^{(k+1)} + Z_k^{(2)} \qquad \mbox{for $k < K$} \\
&= X_K\tilde{a}_{ps}^{(0)}+  Z_K^{(2)} \qquad \qquad \qquad  \qquad \mbox{for $k = K$} \\&
\end{split}
\end{align}

where, $Z_K^{(2)}$ is given by the following equation:
\begin{align}
\begin{split}
Z_k^{(2)} &= \sum_{x=0}^{b_k-1}\left[\frac{1}{b_k}\sum_{i=0}^{x-1}(1 - \beta_{a})^{i}\beta_{a}(1)\right]
\end{split}
\end{align}

Following expression gives the value of $\tilde{a}_{ps}^{(0)}$:
\begin{align}
\begin{split}
 \tilde{a}_{ps}^{(0)} &= \sum_{k=0}^{K} X_k\prod_{l=0}^{k-1}Y_l \tilde{a}_{ps}^{(0)} + \sum_{k=0}^{K} Z_k^{(2)}\prod_{l=0}^{k-1}Y_l 
\\&=\frac{\sum_{k=0}^{K} Z_k^{(2)}\prod_{l=0}^{k-1}Y_l}{ 1- \sum_{k=0}^{K} X_k\prod_{l=0}^{k-1}Y_l}
\end{split}
\end{align}
\end{document}